\documentclass[conference]{IEEEtran}
\pdfoutput=1
\IEEEoverridecommandlockouts
\usepackage{textcase}
\usepackage{cite}
\usepackage{amsmath,amssymb,amsfonts}
\usepackage{algorithmic}
\usepackage{graphicx}
\usepackage{textcomp}
\usepackage{array}
\usepackage{xcolor}
\def\BibTeX{{\rm B\kern-.05em{\sc i\kern-.025em b}\kern-.08em
    T\kern-.1667em\lower.7ex\hbox{E}\kern-.125emX}}
\begin{document}

\title{Fuzzy C-Means Clustering and Sonification of HRV Features\\
\thanks{NSERC-CREATE: Complex Dynamics and Muvik Labs}
}

\author{\IEEEauthorblockN{1\textsuperscript{st} Debanjan Borthakur}
\IEEEauthorblockA{\textit{ McMaster University} \\
Hamilton, Canada \\
borthakd@mcmaster.ca}\\\\
\and
\IEEEauthorblockN{2\textsuperscript{nd} Victoria Grace}
\IEEEauthorblockA{\textit{Muvik Labs} \\
New York, USA \\
vic@muviklabs.io} \\\\\\
\and
\IEEEauthorblockN{3\textsuperscript{rd} Paul Batchelor}
\IEEEauthorblockA{\textit{Muvik Labs} \\
New York, USA \\
paul@muviklabs.io }
\and
\IEEEauthorblockN{4\textsuperscript{th} Harishchandra Dubey}
\IEEEauthorblockA{\textit{UT Dallas, TX, USA} \\
\& Microsoft Corporation, Redmond, WA, USA\\
 Harishchandra.dubey@utdallas.edu}
\and
\IEEEauthorblockN{5\textsuperscript{th} Kunal Mankodiya}
\IEEEauthorblockA{\textit{ University of Rhode Island, Kingston, RI} \\
kunalm@uri.edu}
}

\maketitle

\begin{abstract}
Linear and non-linear measures of heart rate variability (HRV) are widely investigated as non-invasive indicators of health. Stress has a profound impact on heart rate, and different meditation techniques have been found to modulate heartbeat rhythm. This paper aims to explore the process of identifying appropriate metrices from HRV analysis for sonification. Sonification is a type of auditory display involving the process of mapping data to acoustic parameters. This work explores the use of auditory display in aiding the analysis of HRV leveraged by unsupervised machine learning techniques. Unsupervised clustering helps select the appropriate features to improve the sonification interpretability. Vocal synthesis sonification techniques are employed to increase comprehension and learnability of the processed data displayed through sound. These analyses are early steps in building a real-time sound-based biofeedback training system.\\
\end{abstract}

\begin{IEEEkeywords}

HRV, Sonification, clustering
\end{IEEEkeywords}

\section{Introduction}
Several previous studies have investigated the importance of heart rate variability (HRV) measures as a non-invasive indicator of the state of health. HRV is a non-invasive indicator that reflects the balance of the autonomic nervous system \cite{b3}. By indicating how much the heart rate varies over time, HRV can imply the health status of the person. Meditation has a profound effect on HRV. Several studies have investigated the effect of breathing on HRV \cite{b4}.
Along with linear measures of HRV, studies have shown that nonlinear measures of HRV are sensitive to changes in breathing pattern but not light exercise \cite{b1}.  Several nonlinear metrices of heart rate variability such as Symbolic analysis, Shannon Entropy, Rnyi Entropy, Approximate Entropy, Sample Entropy, and Detrended Fluctuation Analysis are there. Studies have found significantly higher Detrended Fluctuation Analysis during slow breathing exercises \cite{b2}. They also observed significantly reduced Approximate Entropy during slow breathing. They concluded that nonlinear behaviour is decreased in slow breathing exercises of heart rate dynamics in healthy young men analyzed through detrended Fluctuation Analysis (DFA) and various entropies. Previous studies have also found that rhythmic recitation of mantra and ava maria produces narrower spectral peaks whereas free talking produces broader peaks as compared to spontaneous breathing \cite{b5}. Visual representation of these changes in the HRV and respiration is a standard method for data display. These physiological signals can also be represented through sound, which is a process referred to as sonification. 
Although biofeedback systems typically use graphic displays to present information, auditory displays can have advantages over visual displays because of the high temporal, amplitude, and frequency resolution of the auditory system. Since human hearing is especially adept at detecting  temporal patterns, sonification of biometric data can lead to better mindfulness practice. Viewing images or graphs can be distracting and wearisome for individuals that want to practice meditations. Sound can also improve listeners focus and pleasure during biofeedback training \cite
{b14}. This paper used data collected from individuals practicing different meditation techniques. Meditation and mindfulness embody many similarities. Practicing mindfulness encompasses a number of various approaches that bring user awareness to focus on the present moment. A growing number of mindfulness exercises concentrate on the tradition of focused-attention meditation. Using this technique, the practitioner is led to shift mental attention away from external stresses toward internal sensations such as breathing. The practice of sustained focused attention on breathing has been shown to improve well-being and reduce stress. Perhaps most importantly, this technique cultivates introspective awareness, the ability to receive and attend to the signals that originate within the body. This awareness has been shown to improve attention-task performance and regulation of emotions \cite{b15}. Sonification can be a useful tool for practicing mindfulness or biofeedback training. Visualization of data lacks in its ability to represent data that dynamically changes over time, which makes sonification a suitable complement or alternative.\par
Parameter mapping and model-based mapping of the data to acoustic parameters are common in sonification. Sonification can have multiple utilities. Bio-feedback might also find its use in intelligent health recommend  systems \cite{b19}.
In \cite{b6} authors have investigated the effectiveness of representing HRV data with auditory interfaces as a supplement or complement to visual displays. In this paper, we have adopted a similar approach. We extracted the HRV features from the beat to beat intervals of three data sets, namely, chi, yoga, and normal breathing, and applied Fuzzy C-Means clustering. The clustering of the data gives us a hint of the features that can be useful in classification. Based on the plots we selected one feature for sonification.

\begin{figure*}[t]
\centering
\includegraphics[width=450pt]{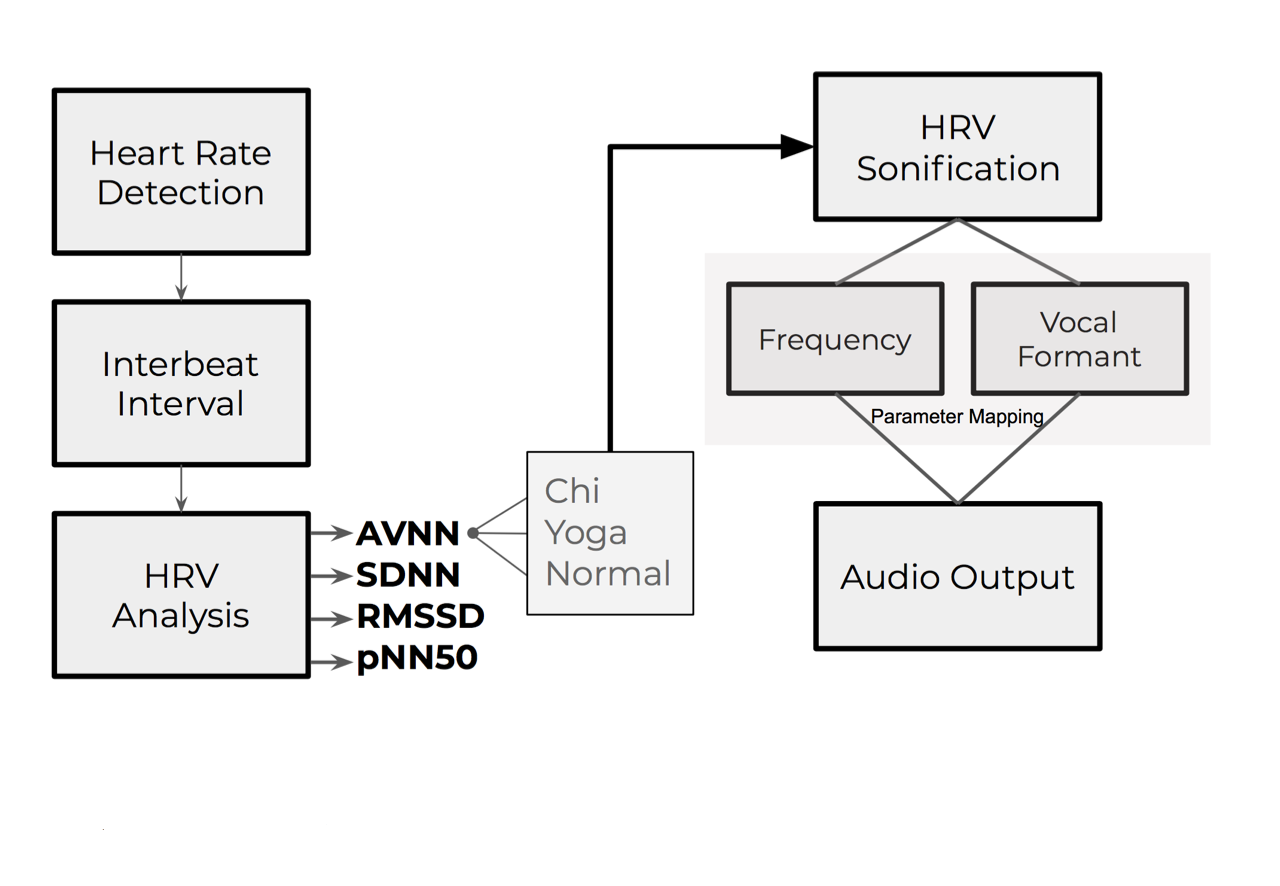}
\caption{Sonification Architecture}
\label{fig1}
\end{figure*}
\section{Related Work}

\subsection{Sonification}
The basic principles of sonification are analogous to visualization. Visual displays employ shapes, lines, and colors, to represent data, while sonification maps data to sonic attributes such as frequency, spectrum, duration, and intensity, and their musical correlates pitch, timbre, rhythm, and volume. Sonification can be used for various types of applications. An auditory display typically represents a data series with a succession of non-speech sounds. Some examples where it has been useful include weather pattern monitoring, volcanic activity detection, and medical and surgical tools. Sonification of heart rate or heart rate variability  can allow for many applications in self-regulation of biometrics. 
 Sonification can map biometric data to different sound dimensions, as well as capture the interactions between these dimensions \cite{b16}.
The sonified sound can be used to provide feedback to the user to assist them in achieving exercise, breathing, and meditation goals.
 In \cite{b7} authors describe Cardiosound, 
which is a portable system to sonify ECG heart rhythm disturbances. Authors
mention three different sonification designs employed. 1) Polarity pitch, 2) ECGrains and 3) Marimba.
 The polarity pitch sonification intends to magnify the time difference between expected and actual
 R peaks in the ECG waveform. In ECGrains a granulated sound is triggered whenever an R peak is found. 
The difference between actual and expected R peaks is considered. The rhythmic pattern between pathological and healthy signals will thus differ based on the grain quantity they will have.
The Marimba sonification is similar to ECGrains. It  uses Marimba sound instead of sine waves. The Breathing pattern can be a useful metric for sonification. In \cite{b8} authors have used a  sound synthesis software to sonify the heart rate variability data. They mapped each cardiac inter beat interval to a pitch sounded by an oscillator that produces short sine wave sounds or grains. They also used inter beat intervals as a clock to provide the timing for each granular event. In \cite{b6} author have examined the categorization task performance of their sonified signals. They found the accuracy of meditation types and states based on their developed auditory interface as 85.6 percent. It shows the effectiveness of sonification. The paper does not directly mention which HRV metrics are sonified and what was the basis of selection of those metrics.
In our work we have used Fuzzy C-Means clustering, which graphically shows the features best at classifying the three different meditation techniques used in this study.
\subsection{Fuzzy clustering}
Fuzzy clustering or soft clustering is a form of  clustering technique where each data point can belong to one or more clusters. This is different to hard clustering. Authors in \cite{b18} discusses K-means clustering, which is a hard clustering method. This technique is developed by J.C Dunn in 1973 and improved by J.C. Bezdek in 1981. The steps involved in Fuzzy C-Means algorithms are listed below:
\begin{itemize}
    \item  Number of clusters are chosen.
    \item  Coefficients are  randomly assigned to the cluster.
    \item Process is repeated until the algorithm converges.
    \item Centroids are calculated for each cluster and coefficients are calculated for each data point.
    \end{itemize}

This algorithms is based on the minimization of the objective function:
\[
J_m=\sum^N_{i=1} \sum^C_{j=1} u_{ij}^m|| X_i - C_j ||^2 \ \ \ \ \ \ \ \rm 
\]
Here m is any real number greater than 1. Here u is the degree of membership of 
X in the cluster, C is the center of the cluster. The norm expresses the similarity of the measured data and the center.
In\cite{b10} authors took a similar approach as discussed in this paper. They applied Fuzzy C-Means clustering and Euclidian distance measure on the statistical features, such as  heart rate, approximate entropy, mean R amplitude, mean R-R interval, standard deviation of normal to normal R-R intervals (SDNN) and root mean square of successive heartbeat interval differences (RMSSD).

\subsection{Data Collection}
The dataset is collected from physionet.org.
Dataset consists of 1) Chi meditation group (age range 26 to 35, mean 29 yrs), 2) Kundalini Yoga meditation group (age range 20 to 52, mean 33 yrs), 3) Spontaneous breathing group (age range 20 to 35, mean 29). The Chi mediator were all graduate and post-doctoral students. The Kundalini Yoga subjects were  at an advanced level of meditation training as mentioned in \cite{b9}. We select 4 subjects' data from each group for the meditation period for analysis. So total 4*3=12 subjects data is analyzed in this paper. The authors in \cite{b9} quantified the heart rate dynamics using the dataset. They observed prominent heart rate oscillations, which are associated with slow breathing.

\subsection{Data Analysis}

Raw heart rate data is collected from physionet. Figure \ref{fig1} shows the sonification architecture. A Matlab based toolbox mhrv is used for calculation of heart rate variability metric from the inter-beat intervals \cite{b11}. We have used the function mhrv.hrv.hrv\_time to get the HRV metrices. They consist of the following:  1) Average of all NN intervals (AVNN, SDNN, RMSSD), and 2) The percentage of NN intervals which differ by at least x (ms)  from their preceding interval (pNNX) respectively. The results are z score normalized prior to clustering by Fuzzy C-Means algorithm. We have used Matlab's fcm function for clustering:
[centers,U] = fcm(data,Nc) , for clustering the heart rate variability metrices that are calculated using the mhrv toolbox. Here centers is the center of the clusters, U is the fuzzy partition matrix. Nc is the number of clusters.

\begin{figure}[t]
\centering
\includegraphics[width=250pt]{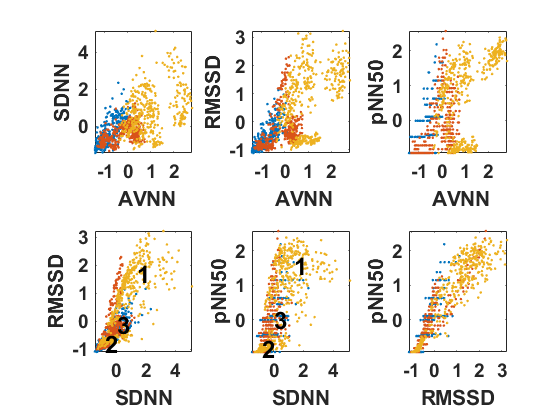}
\caption{Fuzzy -C Means  Clustering}
\label{fig2}
\end{figure}
\begin{figure}[t]
\centering
\includegraphics[width=250pt]{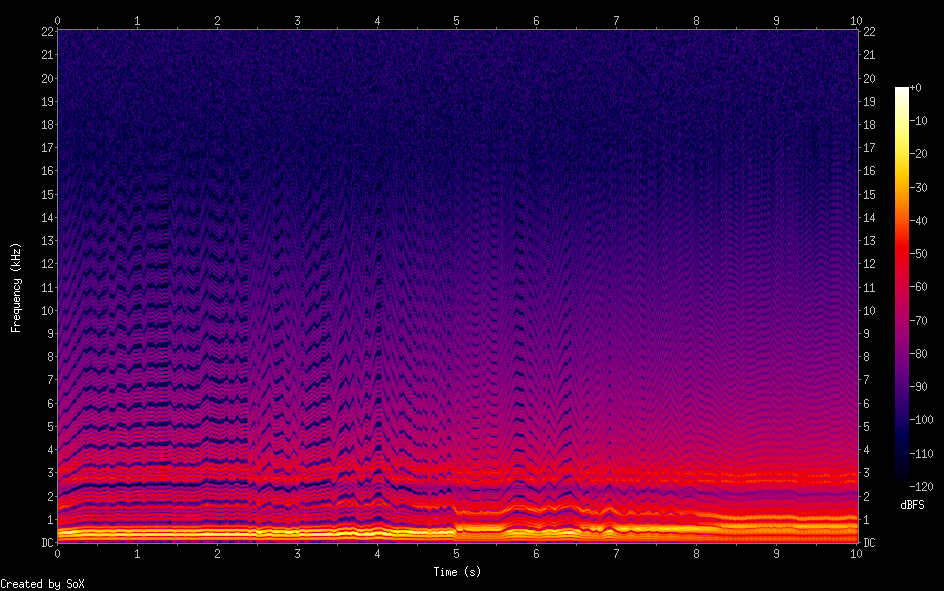}
\caption{Spectrogram of the sonified features of  Meditation Techniques Chi}
\label{figx}
\end{figure}
\begin{figure}[t]
\centering
\includegraphics[width=250pt]{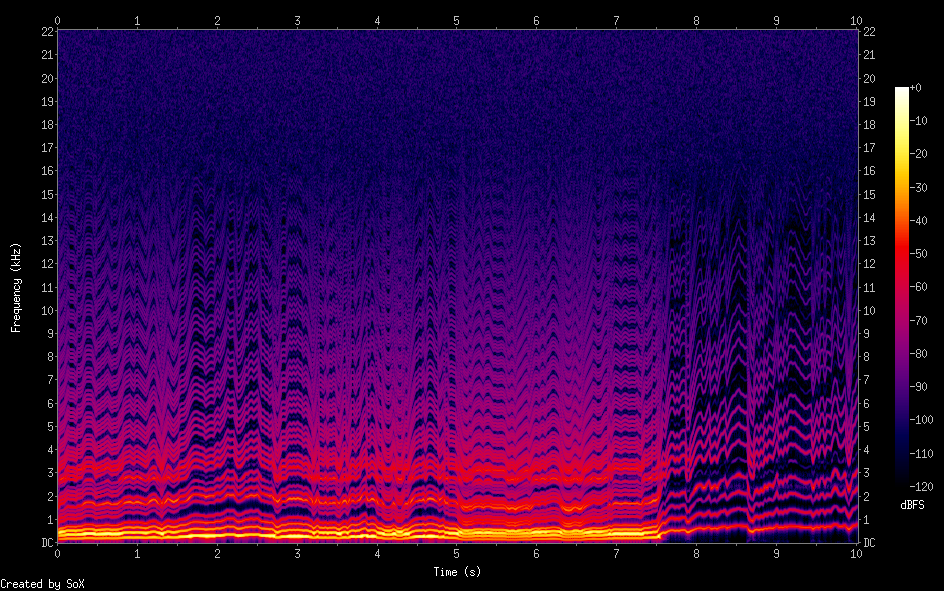}
\caption{Spectrogram of the sonified features of  Normal breathing}
\label{figy}
\end{figure}
\begin{figure}[t]
\centering
\includegraphics[width=250pt]{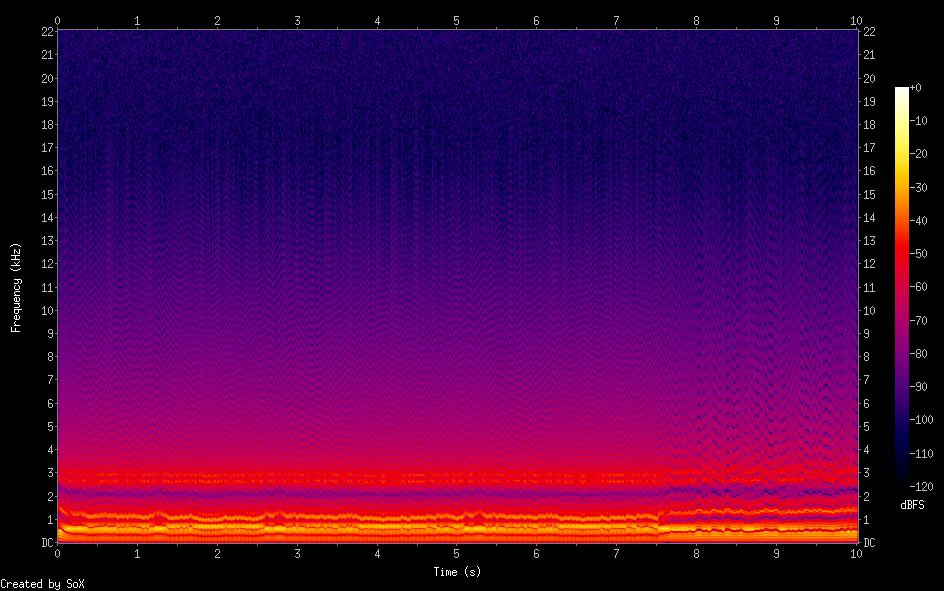}
\caption{Spectrogram of the sonified features of  Meditation Techniques Yoga}
\label{figz}
\end{figure}

Figure \ref{fig2} shows the Fuzzy C-Means clustering plot for HRV features. The data is plotted as combination of two dimensions. The parameters for the Fuzzy C-Means are chosen as number of clusters=3, Fuzzy partition matrix overlap= 2.0, Maximum number of iterations=100. In each iteration the fcm calculates the cluster centers and  updates the fuzzy partition matrix and then computes the objective function value. The clustering ends when the objective function is not improved further. We have considered six combinations of features, which are, SDNN vs AVNN, RMSSD vs AVNN, pNN50 vs AVNN, RMSSD vs SDNN, pNN50 vs SDNN, pNN50 vs RMSSD. 
The next step was the sonification of the HRV metrics. We initially used 'sonify' function in R \cite{b12} to represent the features of HRV through sound, which involved a simple mapping of the normalized data values to pitch. Then we tried a different sonification technique involving formant synthesis using an audio synthesis engine developed by Muvik Labs. The sampling rate of the sound is 44100 Hz. AVNN features from HRV analysis were sonified using a combination of frequency mapping reinforced with a form of formant synthesis. The formant synthesis approach utilizes a bandlimited narrow pulse wave put through a bank of four Butterworth bandpass filters in series, tuned to approximate the first four formants of a tenor voice \cite{b17}. The data mappings control an alpha parameter that linearly interpolates between two filter bank states, shifting between a tenor 'a' and a tenor 'i' vowel sound. The intuition behind choosing to use formant synthesis was that it could potentially activate the linguistic centers of the brain, making it easier to comprehend and recall patterns in the sound. Instead of 'sonify' function of R, we have chosen the aforesaid method for improved intelligibility. The figures  \ref{figx},\ref{figy},\ref{figz} shows the spectrogram of the sonified HRV features of Chi meditation technique, Normal breathing technique, and Yoga meditation technique respectively.

\section{Results and Discussions}
Fuzzy C-Means is effective in showing visual difference between classes \cite{b10}. The three classes that we have used were Chi meditation group, Kundalini Yoga meditation group and Spontaneous breathing group. The dataset was converted to RR intervals for analysis. The RR interval is the interbeat interval. For the HRV analysis we restricted ourselves to the time domain analysis only. Study \cite{b13} has found changes in relative very low frequency (VLF) power before and during exercise. Spectral domain features can also be calculated by the toolbox we are using, but we restricted this analysis to time domain analysis as the goal is not to compare different metrices in the three different meditation techniques. Rather we aimed at clustering the data pairwise to investigate the features which are best at classifying or clustering the three groups. \par
In Figure \ref{fig2} the pairwise FCM clustering plots are shown.  FCM clustering plots of SDNN vs AVNN and RMSSD vs AVNN shows three distinct clusters accurately. Visual inspection of other pairs such as pNN50 vs AVNN and pNN50 vs SDNN does not show distinct clusters. The Euclidean distance measurement of each pair of feature center points will provide more quantified assessment which is not shown in this work. The centers of  each of the clusters for all four HRV features are shown in the Table I. The rows in the table are for different cluster centers. \par
The next phase of the work was the sonification of the HRV features. In \cite{b6} authors sonified the HRV features using a linear mapping. Although which time and frequency domain features were sonified was not explicitly mentioned. They also investigated the learnability, confidence, performance and latency in the classification of sonified data by subjects. We did not yet evaluate the sonified output of the HRV features. This will be a future step that will be taken. For sonification we have chosen the HRV feature metric AVNN. Figures \ref{figx}, \ref{figy}, \ref{figz} are the spectrogram of the sonified signals for the three meditation techniques respectively. 

\begin{table}[h!]
\centering
\begin{tabular}{|c c c c c |} 
 \hline
 Cluster & AVNN & STDNN & RMSSD & pNN50 \\ [1ex] 
 \hline
Cluster1 & 1.30863 & 1.37284 & 1.7099 & 1.57552 \\ [1ex]  
Cluster2 & -0.88163 & -0.76224 & -0.77548 & -0.82544  \\[1ex]  
Cluster3 & 0.15849 & 0.03227 & -0.10038 & 0.01418\\[1ex]  
\hline 
 \\
\end{tabular}
\caption{Cluster centers for all the features} 
\label{table:1}
\end{table}
SoX, a command line sound processing tool, was used to generate the spectrogram plots with the Discrete Fourier Transform (DFT) algorithm. Each sonification WAV file was rendered into a Portable Network Graphic (PNG) file showing time in the X-axis, frequency in the Y-axis, and audio signal magnitude in the Z-axis represented by the colour \cite{b18}. Visual inspection of the spectrograms reveals the distinctive nature of the sonified sounds. An evaluation done by \cite{b6} can be helpful in validating the effectiveness of the sonified signals. We adpot a formant synthesis sonification approach with the intuition that it could potentially activate the linguistic centers of the brain, which will make it easy to learn and recall the sound better. Although no quantitative measures have been taken yet to evaluate the efficiency of this sonification technique, an initial validation of the technique's appropriateness was provided from verbal feedback of four individuals (two musicians and two non-musicians). A simple A-B test compared individual responses to the simple sonifcation technique, mapping the normalized data to frequency (using R toolbox), with the multi-dimensional vocal synthesis technique (using Muvik Labs’ audio synthesis engine). Four out of four individuals reported they found the vocal synthesis sonifications more interesting, and easily memorable. The vocal-like sonifications evoked noticeably more excitement in two of the listeners, who attempted to imitate the fluctuating sonification sound. This response was not perceived after the playing the simple frequency mapping sonification technique.  

\section{Conclusion}
Sonification has more extended scopes as an effective method of biofeedback training. In this work, we have clustered the HRV metrices and have shown the potential of sonifying the HRV features. It is also visible from the plots that FCM clustering can reveal distinct differences between the classes. It can have applications in classifying different types of stress, and the effects of different mindfulness exercises for managing reactions to stress. The sonification of the relevant metric can help provide real-time feedback to a person while performing an exercise in a guided manner. As future work, we should look for other techniques to effectively identify the appropriate HRV features to be used for sonification. Additionally, we plan to expand on the formant synthesis technique for sonification by exploring different mappings of sonic attributes in the future.

\section*{Acknowledgment}
We thank NSERC-CREATE: Complex Dynamics and Muvik labs or supporting this  research.

\vspace{12pt}

\end{document}